\documentclass[conference,comsoc]{IEEEtran}

\IEEEoverridecommandlockouts

\usepackage[T1]{fontenc}

\usepackage{fancyhdr}
\usepackage{graphicx}
\usepackage{cite}

\ifCLASSINFOpdf
 
\else
  
\fi

\usepackage{amsmath}

\usepackage[cmintegrals]{newtxmath}
\usepackage{dblfloatfix}

\usepackage{setspace}
\usepackage{fancyhdr}
\begin{document}

\title{Memory-aware Online Compression of CAN Bus Data for Future Vehicular Systems 
\thanks{This work was supported in part by the SCALE-IoT Project (Grant No. 7026-00042B) granted by the Danish Council for Independent Research, the Aarhus Universitets Forskningsfond Starting Grant Project AUFF- 2017-FLS-7-1.}
}


\author{
  \IEEEauthorblockN{Niloofar Yazdani, Lars Nielsen, and Daniel E. Lucani}
  \IEEEauthorblockA{DIGIT and Department of Engineering, Aarhus University, Denmark\\
  \{n.yazdani, lani, daniel.lucani\}@eng.au.dk \vspace{-0.5cm} }
}


\maketitle
\thispagestyle{fancy}  
\newtheorem{example}{Example}

\begin{abstract}
  Vehicles generate a large amount of data from their internal
  sensors.  This data is not only useful for a vehicle's proper
  operation, but it provides car manufacturers with the ability to
  optimize performance of individual vehicles and companies with
  fleets of vehicles (e.g., trucks, taxis, tractors) to optimize their
  operations to reduce fuel costs and plan repairs.  This paper
  proposes algorithms to compress CAN bus data, specifically, packaged
  as MDF4 files. In particular, we propose lightweight, online and
  configurable compression algorithms that allow limited devices to
  choose the amount of RAM and Flash allocated to them.  We show that
  our proposals can outperform LZW for the same RAM footprint, and can
  even deliver comparable or better performance to DEFLATE under the same
  RAM limitations.
\end{abstract}


\IEEEpeerreviewmaketitle

\section{Introduction}

Modern vehicles have a large number of sensors, actuators, and
Electronic Control Units~(ECUs) measuring and controlling the
vehicle's functions.  Typically, they communicate using the Controller
Area Network~(CAN) protocol~\cite{specification1991version} due to its
robustness and (relatively) high speed.  In fact, the number of ECUs
and sensors per vehicle nowadays can reach up to 80 and 100,
respectively.  These numbers are expected to grow rapidly to enable
more vehicle functions to become automated (get \emph{smarter}), thus,
generating more and more data~\cite{DemandAuto},
\cite{remeli2019automatic}.  Autonomous and self-driving vehicles are
expected to further increase the number and type of sensors necessary
for their correct operation.

Aside from its primary use (e.g., vehicle control), this data can be
stored or transmitted for analysis and usage by a number of stake
holders, including, car manufacturers, insurance companies, and fleet
operators (e.g., trucks, tractors, taxis) optimizing fuel consumption,
route planning, and carrying out failure diagnostics for planned
repairs.
Amount of data generated for further analysis brings with it critical
challenges for its storage and transmission over communication
networks.

\begin{figure}
	\centering
	\includegraphics[width=0.3\textwidth]{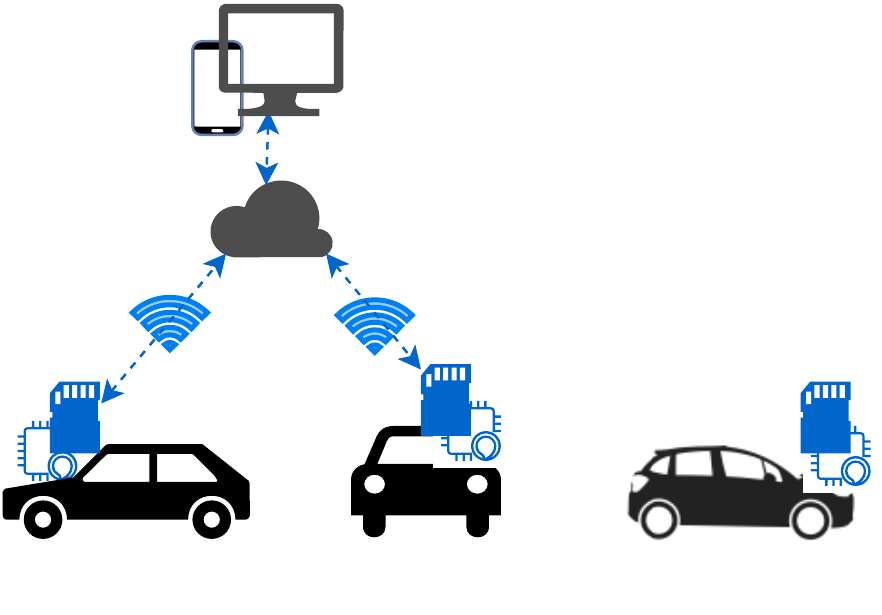}
	\vspace{-7 mm}
	\caption{CAN bus data loggers transferring data to a Cloud.}
	\label{fig:canedge}
\end{figure}

In this paper, we address this problem by proposing a family of
lossless compression schemes based on the emerging concept of
Generalized Deduplication~(GD)~\cite{vestergaard2019generalized}.
The proposed compression algorithms are well suited for vehicular
applications where
(i) perfect reconstruction of the original data is required;
(ii)the vehicle can record data, e.g., CAN bus loggers connected to OBD-II ports that store recordings into an SD card (Fig.~\ref{fig:canedge}), using industry standard file formats, e.g., MDF4~\cite{MDF4};
(iii) the vehicle transmits only when connected to the network sometimes restricted to secure hotspots; 
(iv) data chunks generated by CAN bus are numerous but small;
(v) resource-limited, in-vehicle deployed devices, e.g., a CAN logger, are used for data collection and compression; and
(vi) real-time, online data compression is required to avoid and minimize data loss.


Our techniques achieve these requirements and ultimately (a) reduce the storage requirements and costs at the local device (e.g., logger) and at sink nodes (e.g., Edge/Cloud in Fig.~\ref{fig:canedge}), and (b) reduce the time for transmission from the device to the hotspot/sink node, which allows for a more efficient collection of the data in secure locations (c) allow devices to operate for longer periods with the same persistent storage due to the reduction of overall storage. If network connectivity is continuous, our approach can also provide a mechanism to reduce network costs as our compression can be performed in an online fashion.
In this work, we focus on compressing MDF4 files locally, which requires us to take into account the limited RAM and flash memory available on the local device as part of our design.

Our measurement results show that our techniques perform particularly well under very limited RAM constraints, where standard techniques do not. 
In fact, our techniques can provide compression gains of up to $9.8$ for $1$~kB RAM while DEFLATE requires roughly a minimum of $3$~kB RAM for its operation. 
Even when relaxing RAM constraints, our techniques can provide up to $33$~\% and $75$~\% better compression compared to DEFLATE~\cite{RFC1951-DEFLATE} and LZW~\cite{10.1109/MC.1984.1659158}, respectively.  

\section{Related Work} \label{sec:relatedwork}

Data compression has been extensively researched with many practical compression algorithms.
These fall into two categories: lossy and lossless.
For our scenario, lossy compression is not an option as operators and manufacturers require no distortion or errors as a result of the compression process.
Although there is a wide range of available lossless compression algorithms, e.g., DEFLATE, LZW, LZSS~\cite{10.1145/322344.322346}, some are not suited to operate on resource-limited devices typical in our application. This can be in part due to the large complexity (runtime) of the algorithm or their memory usage. These algorithms may be changed either to fit limited energy resource constraints, e.g., S-LZW ~\cite{10.1145/1182807.1182834}, or computation resource constraints.
Many of these algorithms are usually not developed to carry out compression in an online fashion, i.e., they require all data to be available prior to compression.
String-replacement algorithms,e.g., those based on LZ77~\cite{1055714} and LZ88~\cite{1055934}, are limited to only replacing identical strings.
This means that for data where a single or few bits and bytes changes over time, these algorithms will require larger dictionarie to match each individual (different) sequence/string.    

\section{Background} \label{sec:background}
\begin{figure}
	\centering
	\includegraphics[width=0.22\textwidth]{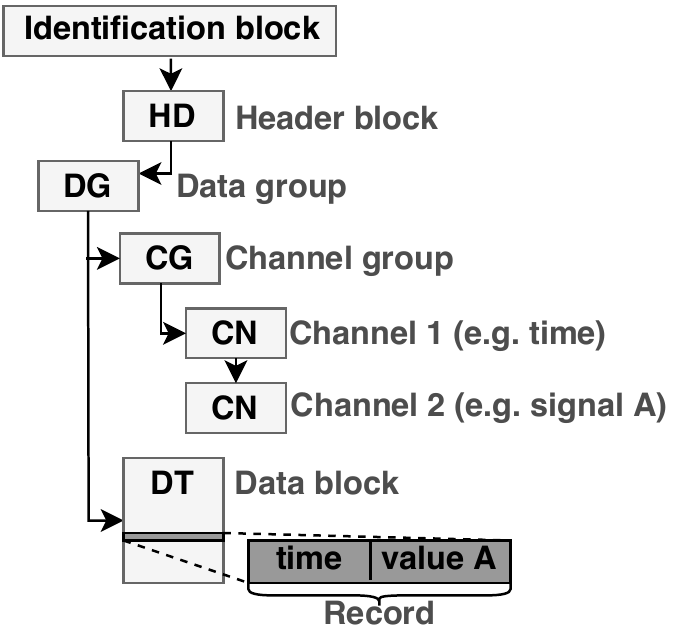}
	\vspace{-4 mm}
	\caption{A block hierarchy for a simple MDF4 file.}
	\label{fig:hiky}
\end{figure}
\begin{figure*}[t]
	\centering
	\includegraphics[width=0.87\textwidth]{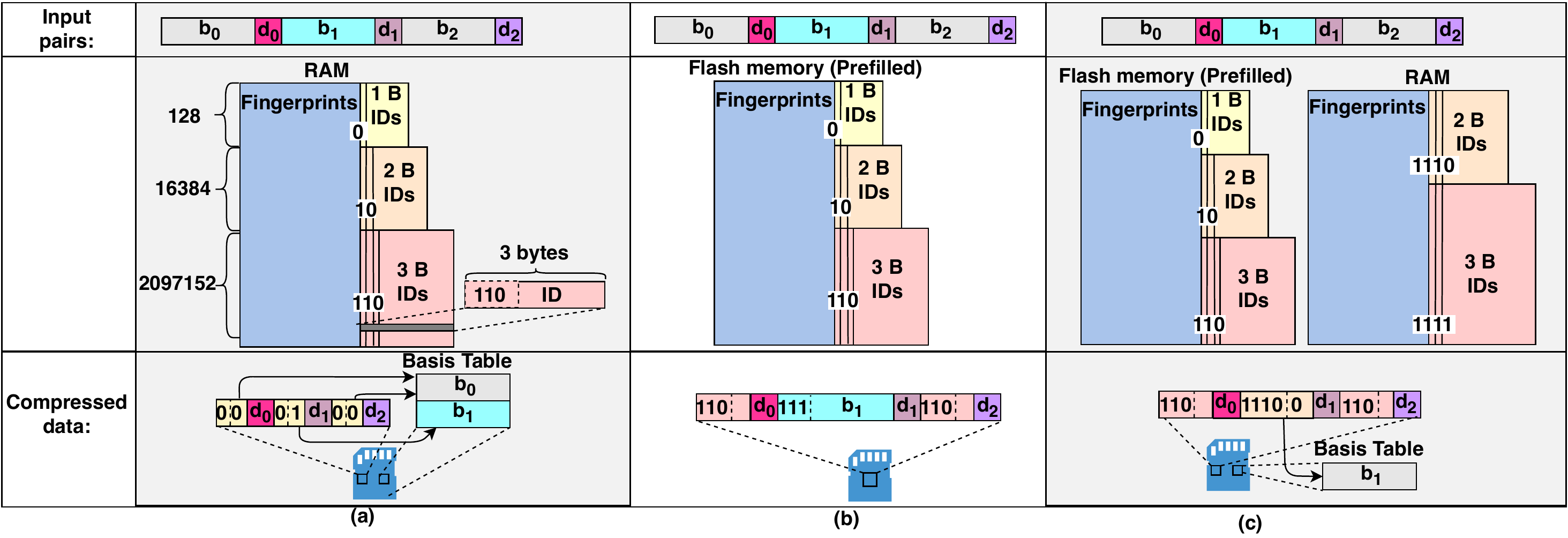}
	\vspace{-4 mm}
	\caption{a) Limited RAM, no flash memory, b) Limited flash memory, no RAM to keep fingerprints, c) Limited flash memory, limited RAM.}
	\label{fig:all}
\end{figure*}

\subsection{Controller Area Network}
CAN is a serial communications protocol~\cite{specification1991version} providing in-vehicle communications among the ECUs and other electronic devices.
CAN's data frame contains up to $8$ bytes of data and its network speed ranges from $40$~kbps to $1$ Mbps.
CAN with Flexible Data-Rate~(CAN-FD)~\cite{canfd} is an enhancement of the CAN protocol.
It supports up to $64$ data bytes per data frame and dual bit rate, i.e., nominal bit rate limited to $1$~Mbps and the data bit rate up to $5$~Mbps.
Both CAN and CAN-FD support $4$ different frame types.
Among them, the \textit{data frame} delivers the actual message transmission/reception of data.
This will be the focus of this paper.

There are $4$ different data frame formats all comprising of $7$ different bit fields: Start of frame, Arbitration field, Control field, Data field, CRC field, Ack field, and End of frame.
In the following, we describe the two fields relevant to this paper.

On the one hand, the Arbitration field includes one or more of (but not limited to) a Base Identifier, Identifier Extension Flag~(IDE), and Identifier Extension depending on the data frame format.
The goal is to indentify the message.
The IDE is $0$ for the base frame format and $1$ for the extended one. 
The Base Identifier is $11$~bits (used for the base frame format) and $29$~bits long (combination of a Base Identifier and Identifier Extension) for the extended frame format represented. 

On the other hand, the Control field comprises of one or more of (but not limited to) Data Length Code~(DLC), Extended Data Length~(EDL), IDE, and Bit Rate Switch~(BRS). 
The DLC indicates the number of bytes in the Data Field, while the EDL distinguishes between CAN or CAN-FD frame formats.
The BRS determines whether the bit rate is switched inside a CAN FD frame format.
The IDE serves a similar purpose as in the Arbitration field.
Finally, the Data Field includes the actual data being transmitted.

\subsection{ASAM MDF file}
ASAM~(Association for Standardization of Automation and Measuring Systems) proposed the MDF4 file format~\cite{MDF4} (file extension of *.mf4) as a binary file format for the automotive industry for recording, e.g., CAN, CAN-FD. 
It holds both raw measurement data and the associated metadata required to interpret it.
An MDF4 file is arranged in binary blocks.
There are various types of blocks abbreviated with $2$ uppercase characters. 
The most relevant block types for this paper are identification block, header block~(HD), data group block~(DG), channel group block~(CG), channel block~(CN), and data block~(DT).
The links among the blocks resemble a tree hierarchy as shown in Fig.~\ref{fig:hiky}.

More specifically, an MDF4 file starts with an identification block to identify the file as a MDF4 file.
This is followed by an HD block as the \textit{root} of the MDF4 file and containing the general description of the file. 
Each DG block holds the list of data groups and refers to plain measurement data, which is stored in \textit{records}.
The records are kept in DT blocks. 
Each record includes the values sampled at the same time, i.e., same time stamp, forming a \emph{row} of values of different types (a tuple). 
The layout of the record is described by the channels~(CN blocks) of the channel group~(CG block).

\subsection{Generalized Deduplication}

Data deduplication~(DD) is a lossless compression scheme which reduces the storage cost by finding equal data chunks and replacing the copies of repeating data chunks by an index/fingerprint.
Recently, generalized deduplication~(GD)\cite{vestergaard2019generalized} was proposed to generalize the concept of DD and reduce the cost of storage.
This is achieved by using a transformation step that allows the system to find equal and similar data chunks for deduplication.
This is done without having to compare to a pool of previously stored chunks to determine similarity.
To achieve this, GD divides each piece of data into a series of equal-size data chunks, $C_i$'s.
Each chunk, $C_i$, is mapped onto a pair of a basis, $b_i$, and a deviation, $d_i$, using a transformation function, e.g., using an error-correcting code~(ECC). 
The basis is large and carries most of the information, while the deviation is normally small and contains information about how to modify the basis to yield the original chunk.
Each basis, $b_i$, is given a fingerprint, $f_{b_i}$, computed using a hash function with a low collision probability (e.g., SHA-1). 
Rather than saving the chunks themselves, a sequence of associated pairs of $(f_{b_i}, d_i)$ is saved.
Then, bases are deduplicated using the same approach as in DD.
Thus, each unique basis will be saved only once. 
Note that DD can be considered as a specific case of GD  where $b_i = C_i$ and there is no deviation.
Recent results in GD showed that it can be used for time-series data and for reducing transmission cost in resource-limited devices~\cite{yazdani2019protocols}.

\textbf{Transformation function:}
The decoding function of an ECC can be used for creating the mapping from a chunk to the associated basis and deviation.
The data chunk is considered as the \textit{codeword}.
By applying the decoding function of the ECC, the \textit{message} is the basis and the deviation carries the information regarding the difference between the codeword and the error-free codeword created by the encoding of the basis.

Hamming codes~\cite{moreira2006essentials} are a family of ECCs capable of correcting one bit error.
Considering $m$ as the number of parity bits, $n=2^m - 1$ is the codeword~(chunk) bit-length. The basis is $k=2^m-1-m$ bits long.
Let us define $H(n,k)$ as a Hamming code with an $n$-bit codeword and a $k$-bit message.
The \textit{syndrome vector} of length $m$ bits indicates which bit is flipped in the codeword compared to the error-free codeword (i.e, the one with a syndrome equal to zero). 
Thus, $m$ bits are sufficient for deviation representation, i.e., representing the changes to the error-free codeword to yield the original codeword/chunk.  

In general, the system's chunk size does not match a Hamming code's $2^m - 1$ bits for requirement for the chunk/codeword for any $m$ value.
A good solution is to use shortened Hamming codes~\cite{macwilliams1977theory}. 
Given $m \geq 3$, any $H^\prime(n-l,k-l)$ forms a shortened 
Hamming code where $l$ is a positive integer and $2^{m-1} < n - l < 2^m - 1$. 
Compared to $H(n,k)$, $H^\prime(n-l,k-l)$ will expect codewords and messages that are $l$ bits shorter for the same parameter $m$.
In essence, this allows a shorter chunk/codeword to be transformed and result in a basis/message that is also shorter by the same amount of bits.
This allows us to avoid inefficiencies coming from zero padding a chunk to fit $H(n,k)$ strict requirements.

\section{Compression Schemes with Limited RAM and Flash Memory}\label{sec:compression}
We focus on compression of MDF4 files' \textit{record}s.
These include all the measurements captured from CAN or CAN-FD at high data rates and some additional information (e.g., timestamps).
These records constitute the majority of the MDF4 file.
Given the device and reliability requirements of vehicular applications, data needs to be compressed at a very high speed considering resource limited devices (i.e., limited RAM and flash memory available, limited processing speed).
Thus, compression algorithms must be lightweight and effective.
In the following, we describe our proposed approach for a new family of compression algorithms based on generalized deduplication.

\subsection{Preprocessing}
Each chunk may contains half, one or more whole rows (records).
For the case half row case, each record is split into two parts and each part forms a chunk. 
If the number of records in each chunk is more than one, then they will appear in order of capture.

\subsection{Compression Algorithm}
By applying GD, each chunk is mapped onto a basis-deviation pair.
Each basis is given a fingerprint or identifier. 
We now describe our proposed compression algorithms using this basis-deviation pairs considering limited RAM to develop a dictionary on-the-fly, limited flash memory to store and use a static dictionary, or by having both of them.

\subsubsection{Limited RAM and No Static Dictionary in Flash }\label{sec:ramnoflash}

A limited RAM is assigned to the algorithm to store a limited number of observed fingerprints. 
The idea is to store as many as possible of the observed fingerprints from the current MDF4 file in RAM to avoid storing a basis twice, as much as possible. 

\textbf{Compression process:}
We keep a table in the RAM as depicted in Fig.~\ref{fig:all}.~(a).
Initially, the table is empty. 
Upon reception, the first chunk is mapped onto a basis-deviation pair.
The compressor searches for the basis' fingerprint in the fingerprint table.
The fingerprint will not match any existing fingerprint because the table is empty.
Then, the compressor saves the fingerprint in the table and assigns a $1$~B ID to it.
The ID starts with \texttt{0} to indicate that it is a $1$~B ID.
For this first basis, the remaining bits in the byte ID are equal to $0$ (first ID saved).
The system saves the ID-deviation pair where the ID is \texttt{00000000} and it saves the basis into a bases table for the current MDF4 file.
The ID-deviation pairs will be stored in the order of the chunks, while the dynamic dictionary in RAM and the bases table will grow only every time a unique basis is inserted (previously bases can be used by multiple chunks). 
The last $7$ bits of the ID show the location of the basis in the bases table.

For the second chunk, the compressor checks for the basis' fingerprint in the table. 
If it matches an existing fingerprint, it only saves the pair of the associated ID and the deviation.
No new fingerprint is kept on RAM, no new ID is generated, and no basis is added to the bases table.
If it does not match to an existing fingerprint, the compressor repeats the process used for the first chunk: saves fingerprint in RAM, assigns the next ID ($00000001$) to it, and saves the basis in the bases table.
Fig.~\ref{fig:all}.~(a) shows an example of storing $3$ chunks using a limited RAM where $b_0 = b_2$. 

This process can be used for the first $128$~($2^7$) unique fingerprints, i.e., they are given a $1$~B ID with a \texttt{0} prefix.
Each of the next $16384$~($2^{14}$) unique fingerprints are given a $2$~B ID starting with a \texttt{10} prefix.
The next $2,097,152$~($2^{21}$) unique fingerprints are given a $3$~B ID starting with a \texttt{110} prefix.
The other steps in the compression process remain unchanged.
Longer IDs can be generated using a similar prefix approach.

The RAM size and the fingerprint length determines the maximum number of fingerprints that can be held simultaneously . 
For example, SHA-1 produces $20$~B hashes.
The first part of the table requires $128 \times (20 + 1) = 2,688$~B.
The second and third parts require up to $360,448$~B and $48,234,496$~B, respectively.


After some point, the allocated RAM for storing fingerprints will be filled.
To maintain a reasonable and adaptive compression, we allow the system to substitute older, less frequently used fingerprints for newer fingerprints.
The main change is that the ID assigned to the new one will always be increasing, i.e., it will not reuse an ID smaller than the latest basis stored.
For example, if the first ever basis is removed from the list the new basis is not going to be assigned the previous ID.
It will be assigned the next ID, which could be of size $2$ or $3$~B.
Due to erasing some of the fingerprints from the RAM, there is a potential to save a basis several times in the bases table.
This does not affect the data recovery, given the ever increasing assignment of IDs, but it may limit compression performance.

\begin{figure}
	\centering
	\includegraphics[width=0.43\textwidth]{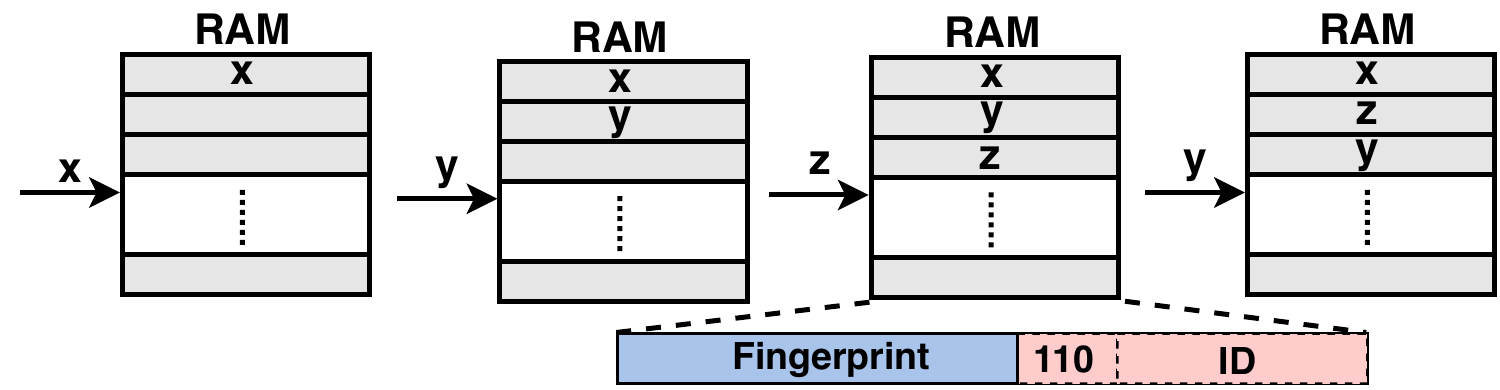}
	\vspace{-4 mm}
	\caption{The idea for keeping the most repetitive fingerprints without counter.}
	\label{fig:ramfill}
\end{figure}

\textbf{Maintaining Most Repetitive Fingerprints in RAM:}
We erase fingerprints which have been used less frequently.
This prevents the compressor from removing IDs linked to basis that are repeated frequently (and provide good compression potential).
Although a counter could be assigned to each fingerprint, this may bias the compressor to fingerprints used heavily for the first records, but which may be used less frequently for records coming in at later stages.
Fig.~\ref{fig:ramfill} describes our approach for updating the bases-ID table in RAM from a logical perspective.
If a fingerprint matches to an already existing fingerprint, we `move' the fingerprint to the bottom of the logical queue.
The top of the queue indicates the first element that will be removed to make room for another.
In Fig.~\ref{fig:ramfill}, fingerprints $x$, $y$ and $z$ are are placed in the first, second and third positions.
However, when fingerprint $y$ is `seen' again, it is moved to the bottom.
Thus, the order changes to $x$, $z$ and $y$. 
If the table becomes full, the compressor easily starts erasing from the top of the table.
In our example, it will erase initially $x$ and $z$ and later $y$. 

\subsubsection{Limited Flash Memory, no Dynamic Dictionary in RAM}\label{sec:flashnoRam}
The flash memory can be used for storing a preset dictionary with the most common fingerprints.
The idea is to have a common dictionary between compressor and decompressor.
Storage cost is reduced as there is no need to build a new bases table, but only to generate a sequence of data containing ID-deviation pairs, for bases whose fingerprint is found in the dictionary, and basis-deviation pairs for bases not included in the dictionary (i.e., basis that need to be sent in full).

  
\textbf{Selecting a Training Set:} 
The dictionary can be created based on already available MDF4 files. 
There are multiple approaches to train the dictionary.
The first one is to use all the available MDF4 files to generate a dictionary, which is common to all the MDF4 files.
This approach allows for the same dictionary to be used for all the vehicles.
A more efficient approach is to have a specific dictionary for any brand and model of vehicles. 
For this, only the MDF4 files related to the same vehicle's brand and model are used to make the dictionary. 
This increases the probability of matching and hence the compression ratio.
A further improvement can be achieved if we train the dictionary for each individual user/vehicle.
The dictionary could be updated frequently for a given vehicle after each travel or even based on intended routes in the future.

\begin{figure}
	\centering
	\includegraphics[width=0.42\textwidth]{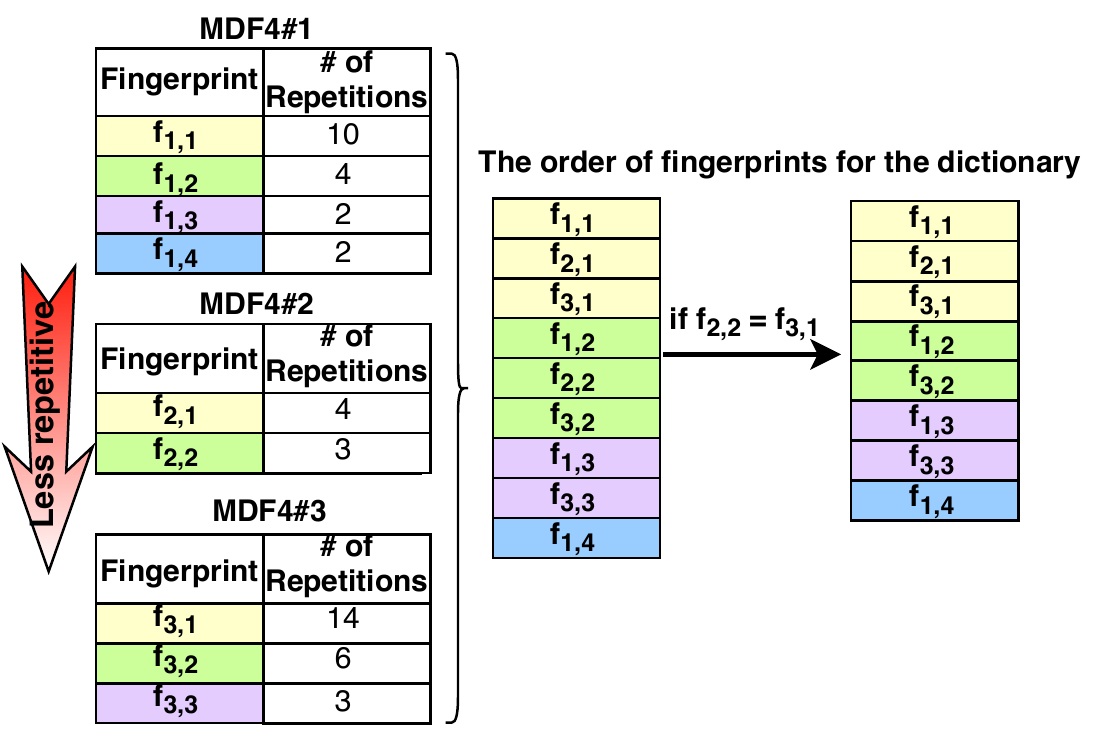}
	\vspace{-4 mm}
	\caption{An example of finding the order of fingerprints for the dictionary.}
	\label{fig:flashorder}
\end{figure}

\textbf{Finding the Most Repetitive Fingerprints:}
After selecting the MDF4 file training set, the next step is to find the most repetitive fingerprints.
Fig.~\ref{fig:flashorder} shows an example of ordering the fingerprints for the dictionary for $3$ MDF4 files.
Fingerprints from each file are organized based on how many times they are repeated.
Then, we pick the most repetitive one from each of them.
Then, we pick the second most repetitive one from each of them, and so on.
Note that in Fig.~\ref{fig:flashorder}, the number of repetitions for the first fingerprint of MDF4\#2 is less than the one for the second fingerprint of MDF4\#3, but, it appears first. 
The purpose is to make sure that the size of a MDF4 file does not affect selecting the fingerprints.
If a fingerprint is already available, we skip it to avoid repetitions.

\textbf{\label{sec:flashonly}Creating the Preset Dictionary:}
Fig.~\ref{fig:all}.~(b) shows the content of the flash memory maintaining a preset dictionary. 
The size of the flash memory and the fingerprint length determines the number of fingerprints in the preset dictionary.
The bases assigned to the fingerprints are kept at the receiver, but not at the compressor to reduce flash usage.
Based on these two factors, a specific number of fingerprints are picked from the list provided in previous subsection starting from the beginning of the list. 
Each of the first $128$~($2^7$) top most repetitive fingerprints are given a $1$~B ID, following the same approach as in Section~\ref{sec:ramnoflash}.
Similarly, we assign the $2$ and $3$~B IDs.


\textbf{Compression using the Dictionary:}
After mapping each chunk onto a basis-deviation pair and assigning a fingerprint to the basis, the compressor look the fingerprint up in the dictionary. 
If available, the compressor only stores the associated ID provided by the dictionary and the deviation.
Otherwise, the whole basis and deviation are stored, prefixed by \texttt{111} to indicate that the following information is a basis.
Fig.~\ref{fig:all}.~(b) depicts the basis-deviation pairs (input) at the top for $3$ consecutive chunks where $b_0$ is equal to $b_2$. Their associated fingerprint is available in the dictionary, resulting in the compressed chunks shown at the bottom.

\subsection{Limited RAM and Limited Flash Memory}
In case of having both a limited flash memory and a limited RAM available, we use the flash memory to keep a preset dictionary as in Sec.~\ref{sec:flashnoRam}.
The IDs associated to the fingerprints in the dictionary are $1$, $2$ and $3$~B long starting with \texttt{0}, \texttt{10}, and \texttt{110}, respectively, as shown in Fig.~\ref{fig:all}~(c).
The application of the RAM is similar to what is described in Sec.~\ref{sec:ramnoflash}. 
However, the IDs are either $2$~B long starting with \texttt{1110} or $3$~B long starting with \texttt{1111} depending on the size of the dictionary.
After applying GD and mapping each chunk onto a basis-deviation pair, the compressor looks up the basis' fingerprint in the flash memory's dictionary. 
If it matches to an existing fingerprint, the compressor saves an ID-deviation pair (similar to Section~\ref{sec:flashnoRam}).
Otherwise, it looks up the basis' fingerprint in RAM and follows the same procedures used to update the fingerprint table in RAM and store ID-deviation pairs and basis persistently from Section~\ref{sec:ramnoflash}.
Fig.~\ref{fig:all}.~(c) shows an example of saving $3$ chunks, where the fingerprint for the first and last one is equal and is available in the flash memory's dictionary.
The second one is added to the dynamic dictionary in RAM.
\section{Performance Evaluation}\label{sec:evaluation}

\subsection{Data Set}
MDF4 files containing CAN or CAN-FD logs are not widely available.
We collected data  using the CAN bus data logger, $CANedge2$~\cite{canedge2} from CSS Electronics,
connected to the OBD-II port of two vehicles: Hyundai Ioniq Plugin Hybrid 2018 and a Toyota CHR 2018.
The device acts as a node on the CAN bus and logs the messages on CAN bus in MDF4 file format.
It logs time series data of in-vehicle sensor information, e.g., vehicle speed, engine speed, fuel consumption rate, into an SD card, with a resolution of up to $50$~us.
Then, it uploads the log files to e.g., a cloud server via WiFi transfer.
We installed firmware version $00.07$~\cite{canedge2}, which was the most up to date at the time. 
The bit rate was selected equal to $500,000$~bps.
The routes were driven both inside and outside the city.
Drivers were free to choose their path. 
A total of $27$ and $33$ different separate measurements were collected for the Toyota and Hyndai vehicles, respectively.
These range from travels of $20$~mins to $120$~mins.
Generated records contain Data field, Identifier, IDE, DLC, EDL, and BRS from CAN, Dir, bus channel, and data length from higher level protocols built on top of CAN, and timestamp. 
Each record is $27$~Bytes.

\begin{figure}
	\centering
	\includegraphics[width=0.45\textwidth]{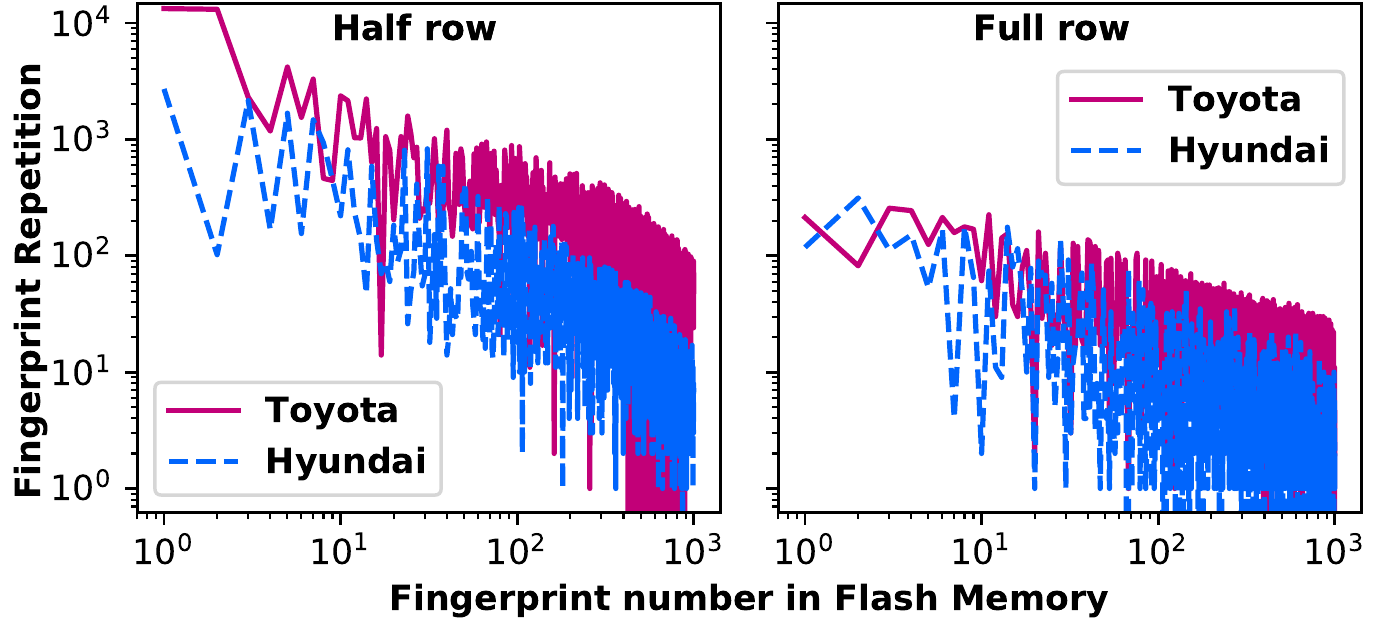}
	\vspace{-4 mm}
	\caption{The average number of repetition of each flash memory's fingerprint.}
	\label{fig:hash-counter}
\end{figure}

\subsection{Numerical Results}
We considered our compression techniques under both GD and the special case of DD.
We use $H^\prime(216,208)$ and $H^\prime(112,105)$ for full row and half row cases, respectively. 
By applying GD, each chunk is mapped onto a basis-deviation pair, while using DD results in a basis equal to the chunk itself and no deviation. 
Each basis is given a fingerprint using CRC32 to maintain a simple fingerprint function at the local device.

\textbf{Bases Similarity Across Data sets:}
A preset dictionary is made for each vehicle's data set, following the approach in Section~\ref{sec:flashonly}.
Fig.~\ref{fig:hash-counter} depicts the average number of repetition of each flash memory's fingerprint using DD for both half and full row in each chunk.
Due to space limitations, we skip the full GD case, but it follows the same trends.
For the half row case, the first fingerprint is repeated $13300$ and $2750$ times on average for Toyota and Hyundai, respectively.
These values are $61$ and $23$ times higher than their counterparts for full row case.
Thus, in the rest of this section we focus on having half a row per chunk.
Our expectation is that even small dictionaries will yield large compression gains in these data sets.

\begin{figure*}
	\centering
	\includegraphics[width=0.82\textwidth]{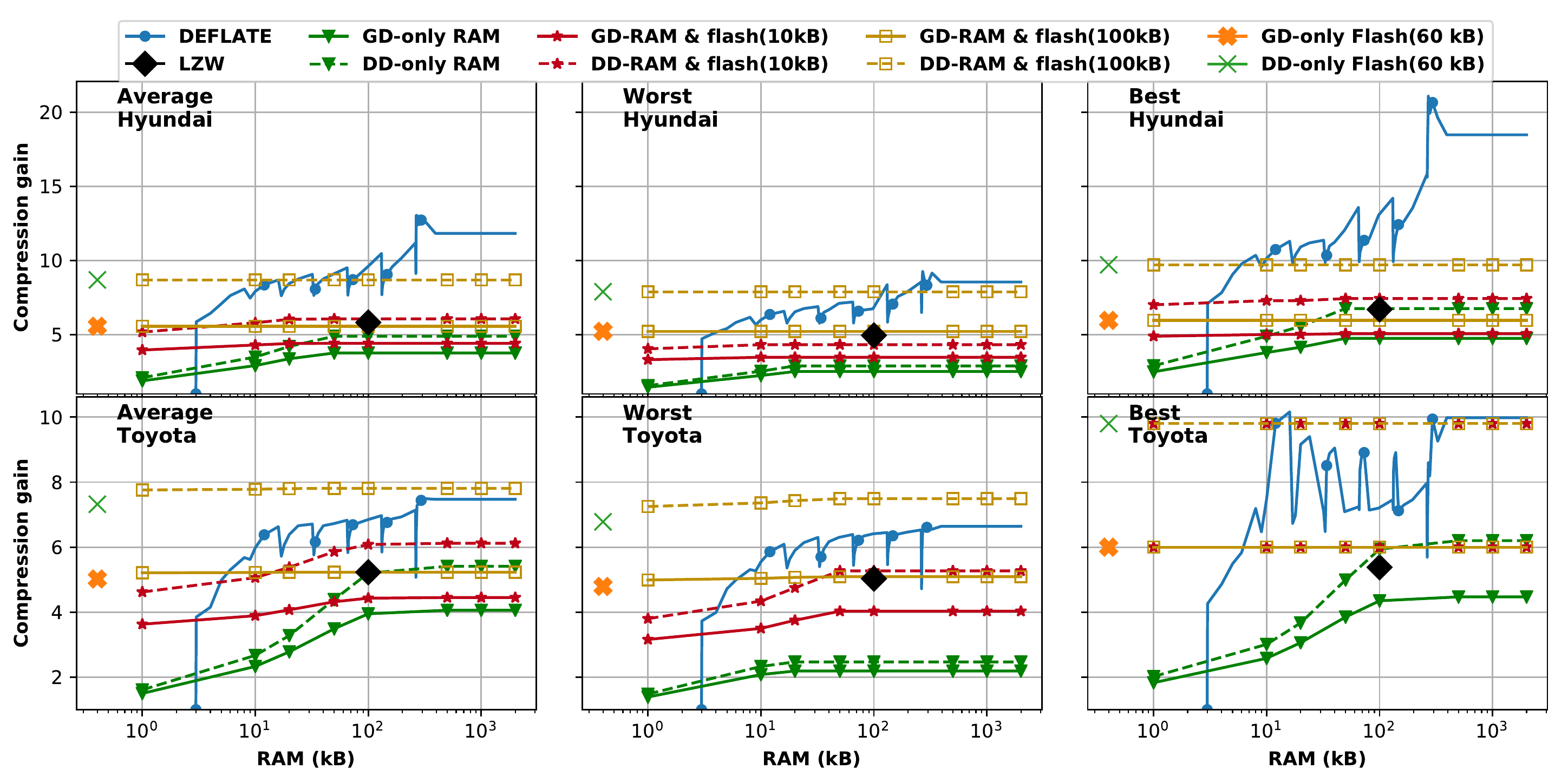}
	\vspace{-5 mm}
	\caption{Average, worst and best compression gains for Hyundai and Toyota data sets under RAM constraints.}
	\label{fig:compression}
	\vspace{-2 mm}
\end{figure*}

\textbf{Compression comparison:}
We consider the compression gain as the uncompressed size divided by the compressed data size, i.e., higher is better.
Fig.~\ref{fig:compression} compares GD and DD in terms of compression gain with LZW and DEFLATE for different RAM constraints.
The average, worst and best compression gains are reported for each vehicle.
We consider differential timestamps for all the compression techniques, i.e., the first row considered has a full representation, the remaining ones are the difference between its own time stamp and the previous row. 
Compression gain for DEFLATE with memory restriction is calculated using python's zlib module and compression for LZW is only reported for the case when there is no RAM limitation and is calculated using python's lzw module. 
Our techniques provide compression gains of up to $10$, i.e., the resulting data is one tenth of the original size.

Fig.~\ref{fig:compression} shows that with $100$~kB RAM  and no static dictionary in flash, our techniques provides an average compression gain of $4.9$ and $5.2$ for Hyundai and Toyota, respectively. This is comparable to LZW's compression gain. 
For case of having only $60$~kB flash memory and no RAM assigned to a keep dynamic dictionary for fingerprints, compression gains of up to $9.8$ are possible for our techniques. 
In fact, with only $10$~kB of flash memory used, our technique outperforms LZW for a wide range of RAM restrictions.
Note that DEFLATE requires a minimum of $3$~kB RAM while our techniques can work with negligible RAM.

Our hybrid approach, using $100$~kB flash memory and limited RAM can outperform both LZW and DEFLATE in most cases. 
In fact, for the case of $20$~kB RAM and $100$~kB flash, our techniques provide up to $8.7$ and $7.8$ average compression gain for Hyundai and Toyota, respectively, while the average compression gain for DEFLATE with the same RAM constraints is $8.45$ and $6.4$, respectively. LZW only provides an average compression of $5.82$ and $5.23$.  
\section{Conclusions} \label{sec:conclusions}
This paper proposes a mechanism for efficient compression of CAN Bus data, particularly, as MDF4 files.
Our proposal allows for online compression, i.e., compressing samples from signals in the CAN bus as they come, that is suitable to provide high reliability and durability in vehicular scenarios.
We show that a judicious and configurable use of available RAM and flash can provide significant compression gains, while also maintaining a simple compression scheme.
In particular, the use of a flash dictionary agreed between sender(s) and receiver (e.g., collecting points, Cloud) can open the door to highly configurable and effective compression in the future.
This is an interesting configuration parameter for service providers to optimize and tailor performance based on their overall system and even individual vehicle characteristics.
Future work will study automated mechanisms to derive the common dictionary and efficiently update it.

As our proposal is based on generalized deduplication, future work will also consider the use of transformations that are tailored specifically for CAN Bus data in order to maximize the potential for compression. We will also consider columnar processing and compression in the future, instead of our current row by row compression strategy.
\section{Acknowledgment}
The authors would like to thank Martin Falch and Christian Steiniche from CCS Electronics Company for providing insight and for supporting us for collecting the data sets.

\ifCLASSOPTIONcaptionsoff
  \newpage
\fi
\bibliographystyle{IEEEtran}
\bibliography{ref}

\end{document}